\documentclass[12pt, fleqn]{article}
\usepackage[cp1251]{inputenc}
\usepackage{latexsym,amsfonts,amssymb}

\newtheorem{theo}{Theorem}
\newcommand{\bt}{\begin{theo}}
\newcommand{\et}{\end{theo}}
\newcommand{\bd}{\begin{displaymath}}
\newcommand{\ed}{\end{displaymath}}

\newcommand{\lf}{\left}
\newcommand{\rg}{\right}

\newcommand{\be} {\begin{equation}}
\newcommand{\ee} {\end{equation}}
\newcommand{\ba} {\begin{array}{l}}
\newcommand{\ea} {\end{array}}

\newcommand{\p} {\partial}

\newcommand{\al} {\alpha}
\newcommand{\lbd} {\lambda}

\begin{document}

{\Large \bf New conditional symmetries and exact solutions of\\
nonlinear reaction-diffusion-convection equations. III}
 \medskip\\
{\bf Roman Cherniha}\footnote{\small e-mail: cherniha@imath.kiev.ua
} {\bf and  Oleksii Pliukhin}\footnote{\small e-mail:
pliukhin@imath.kiev.ua } \\
{\it  Institute of Mathematics, Ukrainian National Academy
of Sciences\\
 Tereshchenkivs'ka Street 3, Kyiv 01601, Ukraine}

\renewcommand{\abstractname}{}
\begin{abstract}
A complete description of $Q$-conditional symmetries of
reaction-diffusion-convection equation with arbitrary power
nonlinearities is finished. It is shown that  the results obtained
in the first and second parts of this work (see arXiv:
math-ph/0612078 and arXiv: math-ph/0706.0814) cannot be extended on
new power nonlinearities arising in the diffusion and convection
coefficients.

\end{abstract}

\begin{center}
\textbf{1. Introduction.}
\end{center}

In the paper \cite{ch-pliu-2006} (see
\cite{ch-pl-06ar}-\cite{ch-pl-07ar} for details) the complete
description of $Q$-conditional symmetries of
reaction-diffusion-convection equations
\be \label {1}
 {U_t}= \ [U ^{m} U_{x}]_{x}+  \lambda U^{m}
 U_{x}+C(U),\ee
 \be \label{1b} U_t=[U^mU_x]_x+\lbd
U^{m+1}U_x+C(U),\ee where $\lambda$ and $ m$ are arbitrary constants
and $C(U)$ is an arbitrary smooth function, has been done. The
symmetries obtained for constructing exact solutions of the relevant
equations have been successfully applied. In the particular case,
new  exact solutions of nonlinear reaction-diffusion-convection
(RDC) equations arising in applications have been found.

The most general RDC equation with power
functions arising in the diffusion and convection coefficients
 reads as
 \be \label {b100}
 {U_t}= \ [U ^{m} U_{x}]_{x}+  \lambda U^{n}U_{x}+C(U).\ee
 In the cases $n=m$ and $n=m+1 $, this equation
coincides with (\ref{1}) and (\ref{1b}), respectively. Here we
report the main result concerning the structure of $Q$-conditional
symmetries of equation (\ref{b100}). Note that equation (\ref{b100})
with $n=0$  is reducing by local substitution $y=x+\lambda t$ to the
equation
\[U_t=(U^mU_y)_y+C(U).\]
$Q$-conditional symmetries of this reaction-diffusion equation  were
investigated in \cite{a-h}, so that we assume $n\ne0$.

\newpage

\begin{center}
\textbf{2. Main Result.}
\end{center}


This is well-known that a (1+1)-dimensional evolution equation may
admit $Q$-conditional symmetries of two different  forms
 \be \label{2}
Q=
\partial_t+\xi(t,x,U)\partial_x+\eta(t,x,U)\partial_U, \ee
and \be \label{2*} Q= \partial_x+\eta(t,x,U)\partial_U,\ee
 where
$\xi$ and $\eta$ are unknown smooth functions, which should be
found.


We have proved that equation (\ref{b100}) with  $\lambda n\ne0,\
n\ne m,m+1 $ cannot   admit any new $Q$-conditional operators of the
form (\ref{2}).
 In other words, a RDC equation with power coefficients
of diffusion and convection admits $Q$-conditional symmetries  of
the form (\ref{2}) only in the cases presented in
\cite{ch-pliu-2006}.

\begin{theo}\label{tab}
The RDC equation (\ref{b100}) with  $\lambda n\ne0,\ n\ne m,m+1 $
 is
invariant only with respect to  operators of the form (\ref{2}),
which are equivalent to a linear combination of the Lie symmetry
operators listed in table 1 of \cite{ch-se-06}.
\end{theo}

It should be stressed that we didn't consider the problem of
constructing $Q$-conditional symmetries of the form (\ref{2*})
because one is equivalent (up to the known non-local transformation)
to solving the given equation (\ref{b100}) \cite{zh-lahno98}.
Obviously, the nonlinear RDC  equation (\ref{b100}) is not
integrable therefore a complete description  of $Q$-conditional
symmetries of this form cannot be derived. On the other hand, one
can try to find {\it particular} solutions of the relevant
determining equation for the function $\eta$, which was derived in
\cite{ch-se-98}, and to construct some operators of the form
(\ref{2*}).

To prove theorem 1   we use the substitution \cite{ch-pl-06ar}
\be\label{b12}\medskip V= \begin{cases}{{U^{m+1},\ m\neq-1,} \cr
{\ln U,\ m=-1.}}\end{cases} \ee In the case $ m\neq-1$ substitution
(\ref{b12}) reduces  equation (\ref{b100}) to the
form\be\label{b101} V_{xx}=V^pV_{t}-\lambda V^k V_{x}+F(V),\
\lambda\ne0;\ p\ne-1;\ k\ne0,p,p+1,\ee
 where $p=-\frac{m}{m+1},\
k=\frac{n-m}{m+1},\ F(V)=-(m+1)C\left(V^{\frac{1}{m+1}}\right),$ \
and in the case $m=-1$ to the form \be \label{b102}
V_{xx}=e^VV_{t}-\lambda e^{(n+1)V} V_{x}+F(V),\ \lambda\ne0,\
n\ne0,-1\ee where $F(V)=-C(e^V)$. We use the work \cite{ch-se-98} to
obtain the system of determining equations for finding the
coefficients of the operator \be\label{b15} Q= \
\partial_t+\xi(t,x,V)\partial_x+\eta(t,x,V)\partial_V,\ee which is
locally equivalent to the operator (\ref{2}) (up to notations).

In the case $F_0(V)=V^p,\ F_1(V)=-\lambda V^k,\ F_2(V)=F(V),$ system
(2.38) \cite{ch-se-98} takes the form
 \be\label{b103}\ba\medskip
\xi_{VV}=0,\\ \medskip\eta_{VV}=2\xi_{V}(-\lambda V^k-\xi V^p)+2\xi_{xV},\\
(2\xi_{V}\eta-2\xi\xi_{x}-\xi_{t})V^{p}-\xi\eta
pV^{p-1}-\lambda\xi_{x}V^k-\lambda k\eta V^{k-1}+\\
\medskip+3\xi_{V}F-2\eta_{xV}+\xi_{xx}=0,\\\eta
F_{V}+(2\xi_{x}-\eta_{V})F+(2\xi_{x}\eta+\eta_{t})V^p+p\eta^{2}V^{p-1}-\\-\lambda\eta_x
V^k-\eta_{xx}=0,\ea \ee and in the case $F_0(V)=e^V,\
F_1(V)=-\lambda e^{(n+1)V},\ F_2(V)=F(V),$ one takes the form
\begin{equation}\label{b104}\ba\medskip \xi_{VV}=0,\\
\medskip
\eta_{VV}=2\xi_{V}(-\lambda e^{(n+1)V}-\xi e^V)+2\xi_{xV},\\
\bigr(\xi_{t}+2\xi\xi_{x}+(\xi+\lambda(n+1)-2\xi_V)\eta\bigr)e^V+\\
\medskip+\lambda\xi_x e^{(n+1)V}
-3\xi_{V}F+2\eta_{xV}-\xi_{xx}=0,\\
\eta F_{V}+(2\xi_{x}-\eta_{V})F+\eta^{2}e^{V}+2\xi_{x}\eta
e^{V}+\eta_{t}e^{V}-\\-\lambda\eta_{x} e^{(n+1)V}-\eta_{xx}=0.\ea
\end{equation}
We divide the solving systems (\ref{b103}) and (\ref{b104}) on three
cases:
\[(a)\ \xi=aV +b,\ \eta=\eta(V),\ a=const, b=const \]
\[(b)\ \xi=a(t,x)V+f(t,x),\ a(t,x)\ne0,\]
\[(c)\ \xi=f(t,x),\ \eta=g(t,x)V+h(t,x).\]One can easily check that these  cases take into account all
possible solutions of the systems (\ref{b103}) and (\ref{b104}).

\begin{theo}\label{tab2}
In the cases $(a)$ and $(b)$ equations (\ref{b101}) and (\ref{b102})
can be invariant only with respect to the   Lie symmetry operator of
the form \be\label{b109} Q=\partial_t+\xi\partial_x,\ \xi=const.\ee
\end{theo}

\begin{theo}
In the case $(c)$ equations (\ref{b101})  and (\ref{b102}) are
invariant only with respect to  operators of the form (\ref{b15}),
which are equivalent to the Lie symmetry operators listed in
\cite{ch-se-98} and \cite{ch-se-06}.
\end{theo}

Theorem 1 immediatelly follows from theorems 2 and 3 if one takes into account that
the RDC equation  (\ref{b100}) is locally equivalent to the equations (\ref{b101}) (if $ m\neq-1$) and (\ref{b102}) (if $ m=-1$).

\newpage
\begin{center}
\textbf{3. Proof of  Theorem 2} \end{center}

Firstly, let us consider the case $(a)$. In the quite similar way as
it was done in  \cite{ch-pl-06ar} (see pp. 11--14) one proves that
operator (\ref{2}) may take only form (\ref{b109}), which is the Lie
symmetry operator of the equations  (\ref{b101}) and (\ref{b102}).

Case $(b)$. Let us consider the system (\ref{b103}) (the
consideration of the system (\ref{b104}) is quite similar). The
general solution of the first equation arising in (\ref{b103}) is
\be\label{b94}\xi=a(t,x)V+f(t,x),\ee the solution of the second
equation of (\ref{b103}) is \be\label{b105}\ba\medskip
\eta=-\frac{2a^2}{(p+2)(p+3)}V^{p+3}-\frac{2af}{(p+1)(p+2)}V^{p+2}-\frac{2\lambda
a}{(k+1)(k+2)}V^{k+2}+\\ \qquad+a_xV^2+g(t,x)V+h(t,x),\ea\ee if
$p\ne-2,-3,\ k\ne-1,-2$ (the consideration of the cases $p=-2,-3,\
k=-1,-2$ is much more simpler). Substituting (\ref{b105}) into the
third equation of system (\ref{b103}) we obtain
\begin{eqnarray}\label{b106}
F(V)=\nonumber&&\frac{1}{3}\Biggr(-\frac{2a^2(p-2)}{(p+2)(p+3)}V^{2p+3}-\frac{2pf^2}{(p+1)(p+2)}V^{2p+1}-\\&&\nonumber-2\lambda
a\left(\frac{k}{(p+2)(p+3)}+\frac{p-2}{(k+1)(k+2)}\right)V^{p+k+2}-\\&&\nonumber-2\lambda
f\left(\frac{k}{(p+1)(p+2)}+\frac{p}{(k+1)(k+2)}\right)V^{k+p+1}-
\\&&\nonumber-4af\frac{p^2+p-3}{(p+1)(p+2)(p+3)}V^{2p+2}+a_x\frac{(p-2)(p+4)}{p+2}V^{p+2}+
\\&&\nonumber+\frac{1}{a}\biggr(\Bigr(2(af_x+a_xf)\frac{p-1}{p+1}+a_t+pfa_x+ag(p-2)\Bigr)V^{p+1}+
\\&&\nonumber+(f_t+2ff_x+pfg+ah(p-2))V^p+pfhV^{p-1}+
\\&&\nonumber+\lambda a_x\frac{(k-1)(k+3)}{k+1}V^{k+1}+\lambda(f_x+kg)V^k+\lambda khV^{k-1}+
\\&&+3a_{xx}V+2g_x-f_{xx}\biggr)-\frac{2\lambda^2k}{(k+1)(k+2)}V^{2k+1}\Biggr).\end{eqnarray}
Since the function $F$ depends only on the variable $V$, all
coefficients by different  powers of this variable must be
constants. In the general case, we obtain fifteen equations for
determining the functions $a,\ f,\ g,\ h$. The number of equations
may be shortened for certain values of $k$ and $p$ .

It turns out that the functions $g$ and $ h$ must be constants, if
$a(t,x)=const, \ f(t,x)=const$ therefore we arrive at  the case
$(a)$, which lead only to the Lie operator (\ref{b109}). Let us
prove this statement.

 Expression (\ref{b106}) with  $a=const, \ f=const$, takes the form
\be\label{b117}\ba
F=\displaystyle\frac{1}{3}\biggr(M(V)+\frac{1}{a}\Bigr(ag(p-2)V^{p+1}+(pfg+ah(p-2))V^p+\\
\qquad+pfhV^{p-1}+\lambda kgV^k+\lambda
khV^{k-1}+2g_x\Bigr)\biggr),\ea\ee where
\[\ba M(V)=-\frac{2a^2(p-2)}{(p+2)(p+3)}V^{2p+3}-\frac{2pf^2}{(p+1)(p+2)}V^{2p+1}-\\
\qquad\qquad\ -2\lambda
a\left(\frac{k}{(p+2)(p+3)}+\frac{p-2}{(k+1)(k+2)}\right)V^{p+k+2}-\\
\qquad\qquad\ -2\lambda
f\left(\frac{k}{(p+1)(p+2)}+\frac{p}{(k+1)(k+2)}\right)V^{k+p+1}-
\\ \qquad\qquad\ -4af\frac{p^2+p-3}{(p+1)(p+2)(p+3)}V^{2p+2}-\frac{2\lambda^2k}{(k+1)(k+2)}V^{2k+1}\ea\]
is  the polynomial, which depends only on $V$. Analyzing powers
$p+1,\ p,\ p-1,\ k,\ k-1,\ 0$ in (\ref{b117}), we receive the
conclusion that there are  only five cases, when two and more among
them are equal, namely:
 \be\label{b118}k=p-1,\ k=p+2,\ p=0,\
p=1,\ k=1.\ee We have to investigate also the general case, when
conditions (\ref{b118}) don't take place. In the general case, all
coefficients by the  powers of $V$ must  be constant. Let us
consider the coefficients by the terms  $V^{p+1}:\
\displaystyle\frac{1}{3}g(p-2)=const$ and $V^p:\
\displaystyle\frac{1}{3a}(pfg+ah(p-2))=const.$ Considering
separately two subcases,   $p\ne2$ and $p=2$ one easily obtains
that  $g(t,x)=const,\ h(t,x)=const$, provided  $a(t,x)=const, \
f(t,x)=const$.

Let us consider the case $k=p-1$ from (\ref{b118}) leading to the
powers  $p+1,\ p,\ p-1,\ p-1,\ p-2,\ 0$ of the variable $V$ in
(\ref{b117}). We must separately  consider the subcases $p=2, p=-1,
p=0, p=1$ and $p\ne2, 1, 0, -1$.

The expression (\ref{b117}) with $p=2,\ k=p-1=1$ gets the form
\be\label{b119}
F=\displaystyle\frac{1}{3}\biggr(M(V)+\frac{1}{a}\Bigr(2fgV^2+(2fh+\lambda
g)V+h+2g_x\Bigr)\biggr).\ee Since the right-hand-sight of
(\ref{b119}) cannot depend on $t,\ x,$ we immediately obtain
$g=const$ and $h=const$.

In the case $p\ne2, 1, 0, -1$,\ $k=p-1$  expression (\ref{b117})
takes the form
\[\ba
F=\displaystyle\frac{1}{3}\biggr(M(V)+\frac{1}{a}\Bigr(ag(p-2)V^{p+1}+(pfg+ah(p-2))V^p+\\
\qquad+pfhV^{p-1}+\lambda (p-1)gV^{p-1}+\lambda
(p-1)hV^{p-2}+2g_x\Bigr)\biggr).\ea\]
 Since  coefficients by
$V^{p+1}$ and $ V^p$ must be constant we again obtain
 $g=const$ and
$h=const$.  Subcase $ p=-1$ contradicts to  the condition
 presented
above in (\ref{b101}),  while the subcases  $p=0$ and $ p=1$
 should be considered for arbitrary $k$ (see (\ref{b118})).

We have checked other cases from  (\ref{b118}) and obtained the same
result.  Thus,  the functions $g$ and $ h$  are some constants in
the expression (\ref{b106}) provided   $a=const,\ f=const$.

On the other hand, the functions $a$ and $ f$ are constants if the
powers $2p+3$ and $2p+1$ are not equal to any other power in
(\ref{b106}).

Thus, to prove   theorem \ref{tab2} we must consider  only such
cases, when the powers $2p+3$ and $2p+1$ are equal to other power(s)
in (\ref{b106}). One can easily check that  this happens only in the
following cases: \be\label{b108}\ba
1)\ p=-4,\ 2)\ p=-\frac{3}{2},\ 3)\ p=-\frac{1}{2},\ 4)\ p=0,\ 5)\ p=1,\ 6)\ p=2,\\
7)\ k=p-1,\  8)\ k=p+2,\ 9)\ k=2p,\ 10)\ k=2p+1,\ 11)\ k=2p+2,\\
12)\ k=2p+3,\ 13)\ k=2p+4.\ea\ee

Let us consider the case 1) from (\ref{b108}) in details (all other
cases can be investigated in a quite similar way). 
Consider   fifteen powers of the  variable $V$,
 arising  in (\ref{b106}) with $k=p-1$: $2p+3, 2p+1,
2p+1, 2p, 2p+2, p+2, p+1, p, p-1, p, p-1, p-2, 1, 0, 2p-1.$
Let us
 form the table 1 with those  values of $p$
listed  in the second row, when at least one of the powers
listed in the first row  is equal to $2p+3$ (we don't write down the
power $2p-1$ because  the corresponding coefficient is already
constant).
\begin{center}\label{b107}
Table 1.
\medskip

\begin{tabular}{|c|c|c|c|c|c|c|c|c|c|c|}

\hline
 &$2p+1$&$2p$&$2p+2$&$p+2$&$p+1$&$p$&$p-1$&$p-2$&$1$&$0$ \\
\hline

$2p+3$&-&-&-&$-1$&$-2$&$-3$&$-4$&$-5$&$-1$&$-\frac{3}{2}$ \\
\hline

\end{tabular}
\end{center}

The values $p=-1,-2,-3$ contradict to the conditions presented above
(see (\ref{b101}), (\ref{b105})). Other values $p$ listed in the
second row of table 1 are only subcases of the corresponding  cases
from (\ref{b108}), namely:
 $p=-\frac{3}{2}$  and $p=-4$ are  subcases of cases
 7) and   11), respectively,  $p=-5$ lead to $k=-6$,  so that
  this is a subcase of 8).

Thus, the power $2p+3$ doesn't coincide with any other, so that,
$a=const$ (see the first term in right-hand-side of (\ref{b106})). Otherwise we should put $p=2$ but this is another case from (\ref{b108}). 

To establish that $f=const$ we  analyze  the power $2p+1$. Let us
form the table 2  in the same  way as we built table 1 and taking
into account that $a=const$. We obtain five special values of
parameter $p$.
\begin{center}\label{b107b}
Table 2.
\medskip

\begin{tabular}{|c|c|c|c|c|c|c|c|c|c|c|}

\hline
 &$2p$&$2p+2$&$p+1$&$p$&$p-1$&$p-2$&$0$ \\
\hline

$2p+1$&-&-&$0$&$-1$&$-2$&$-3$&$-\frac{1}{2}$ \\
\hline

\end{tabular}
\end{center}
Again the values $p=0,-1,-2,-3$ contradict to conditions presented
above (see (\ref{b101}), (\ref{b105})). The last value
$p=-\frac{1}{2}$ listed in the second row of table 2 is only a
subcases of  case 3) from (\ref{b108}).
 This means that the power $2p+1$ doesn't coincide with
any other power, hence, $f=const$ (see the second term in
right-hand-side of (\ref{b106})).

Thus, we obtain that  $a=const$ and $ f=const$ in the case 1) from
(\ref{b108}), so that, taking into account the statement proved
above,  we arrive at the case $(a)$.

 All other
cases from (\ref{b108}) have been  analyzed  a similar way and the
same result established, i.e.  $a=const$ and $ f=const$.

\textbf{The proof is now completed.}

\begin{center}
\textbf{4. Sketch of the Proof of  Theorem 3} \end{center}

 Substituting $\xi$ and $\eta$  from the case $(c)$ into
 the third equation
of  (\ref{b103}) (the investigation of system (\ref{b104}) is
analogous), we obtain \be\label{b154}
(2ff_x+f_t+pfg)V^p+pfhV^{p-1}+\lbd(f_x+kg)V^k+\lbd
khV^{k-1}+2g_x-f_{xx}=0.\ee To analyze (\ref{b154})
 one needs to consider only the following special cases:
  $p=0,\ p=1,\  k=p-1,\ k=1 $ and  the general case, when $p$ and $ k$
don't satisfy   these restrictions.

Consider   the case $p=0$ in detail (the next two cases and  the
general case are investigated in the same way). Equation
(\ref{b154}) with $p=0$
 takes
the form
 \be\label{b155}\lbd(f_x+kg)V^k+\lbd
khV^{k-1}+f_t+2ff_x-f_{xx}+2g_x=0.\ee
 Since $k\ne p=0$ and $k\ne
p+1=1,$  one can split  (\ref{b155}) by different
powers of $V$ and obtain
 the system
 \be\label{b156}\ba \lbd kh=0,\\ \lbd(f_x+kg)=0,\\
f_t+2ff_x-f_{xx}+2g_x=0.\ea\ee Taking into account the restrictions
presented above, we obtain \be\label{b157}h=0, f_x=-kg.\ee
Substituting expressions (\ref{b157}) into the  fourth equation of
(\ref{b103}), we arrive at the linear ODE
\be\label{b158}gVF_V-(2k+1)gF=\lbd g_xV^{k+1}+(g_{xx}+2kg^2-g_t)V\ee
with the general solution
 \be\label{b159}F=\lbd_1V^{2k+1}-\frac{\lbd
g_x}{kg}V^{k+1}+\Bigr(\frac{g_t-g_{xx}}{2kg}-g\Bigr)V.\ee Note the
special subcase  $g=0$ immediately  leads to $h=0$ and $f=const$,
therefore the  Lie symmetry (\ref{b109}) is obtained.

Since right-hand-side of (\ref{b159}) cannot  depend on the
independent variables, we obtain
\be\label{b171}\ba\frac{\lbd g_x}{kg}=\lbd_2,\\
\frac{g_t-g_{xx}}{2kg}-g=\lbd_3,\ea\ee where lambda-s are some
constants. The general solution of  (\ref{b171}) is $g=\al(t)$ and
then, using  (\ref{b157}), we have \be\label{b161} f=-k\al(t)
x+\beta(t),\ee where $\al(t),\ \beta(t)$ are to-be-determined
functions. Substituting (\ref{b161}) into the third equation of
(\ref{b156}), we obtain
\be\label{b162}(-k\al_t+2k^2\al^2)x+\beta_t-2k\al\beta=0.\ee
Obviously, equation (\ref{b162}) is equivalent to two ODE equations
with the general solution
\be\label{b163}\al=-\frac{1}{2kt+A_1},\
\beta=\frac{A_2}{2kt+A_1},\ A_i=const,\ i=1,2.\ee

Finally, taking into account (\ref{b157}), (\ref{b159}),
(\ref{b161}) and (\ref{b163}), we have found that equation
\[V_{xx}=V_t-\lbd V^kV_x+\lbd_1 V^{2k+1},\] is invariant
under the operator
\[Q=\p_t+\frac{1}{2kt+A_1}\Bigr((kx+A_2)\p_x-V\p_V\Bigr).\]
Multiplying this operator by $2kt+A_1$ one obtains the operator
\be\label{b163*} X=(2kt+A_1)\p_t+(kx+A_2)\p_x-V\p_V,\ee
 which is nothing else but a linear combination of Lie symmetry operators
 of this equation (see the case
7 of table 1 \cite{ch-se-06}).

Let us consider the case $k=1$, which is special. Equation
(\ref{b154}) with $k=1$ takes the form \be\label{b164}
(2ff_x+f_t+pfg)V^p+pfhV^{p-1}+\lbd(f_x+g)V+\lbd h+2g_x-f_{xx}=0.\ee
We must consider only two subcases $p\ne2$ and $p=2$ (we remind the
reader that the subcases $p=0$ and $p=1$ were considered above). It
is easily  shown that the first of them leads only to Lie
symmetries, since  we can split the equation (\ref{b164}) with
respect to the  four
 different powers of  $V$, i.e., four equations are obtained.
 In the second subcase (\ref{b164}) takes the form
\be\label{b165} (2ff_x+f_t+2fg)V^2+(2fh+\lbd(f_x+g))V+\lbd
h+2g_x-f_{xx}=0\ee and we obtain only three equations
 \be\label
{b166}\ba 2ff_x+f_t+2fg=0,\\2fh+\lbd(f_x+g)=0,\\ \lbd
h+2g_x-f_{xx}=0.\ea\ee So, (\ref{b165}) is the nonlinear system of
three PDEs for three unknown functions.

 We must also take into account   the fourth equation of
 (\ref{b103}) with $p=2$,  namely:
\be\label{b167}\ba(gV+h)F_V+(2f_x-g)F=-(gV+h)(3gV^2+2hV)-\\
-(2f_x-g)(gV+h)V^2-(g_tV+h_t)V^2+\lbd(g_xV+h_x)V+g_{xx}V+h_{xx}.
\ea\ee Analyzing the first term of  (\ref{b167}) we must consider
three subcases, $1)\ g=h=0,\ 2)\ g=0,\ h\ne0,\ 3)\ g\ne0,$  for
solving this equation. It turns out,  subcases 1) and 2) lead the
Lie symmetry (\ref{b109}) (we omit the relevant calculations because
they are rather simple).

Let us  consider the most complicated case 3) $g\ne0$. The general
solution of the equation (\ref{b167}) is \be\label{b173}F=\lbd_3
V^3+\lbd_2V^2+\lbd_1
V+\lbd_0+A\Bigr(V+\frac{h}{g}\Bigr)^{1-\frac{2f_x}{g}},\ A=const.\ee
Here $\lbd_i,\ i=0,1,2,3$ are  some functions on  $f,\ g,\ h$ and
their derivatives  but we omit those expressions  for lambda-s
because they are rather cumbersome. Since the function $F$ must
depend only on $V$, then we obtain
\be\label{b174}1-\frac{2f_x}{g}=A_1^*, \quad A_1^*=const \ee if
$A\ne0$.  Formula (\ref{b174}) may be written in the form
\be\label{b175}g=A_1f_x,\quad A_1=\frac{2}{1-A_1^*} \ee
 (the case
$A_1^*=1$ leads to the contradiction $g=0$). Substituting
(\ref{b175}) into  the third equation of  (\ref{b166})  we obtain
\be\label{b176}h=\frac{1-2A_1}{\lbd}f_{xx}.\ee Substituting
(\ref{b175}) and (\ref{b176}) into the second equation of
(\ref{b166}), we arrive at
\be\label{b179}2(2A_1-1)ff_{xx}=\lbd^2(A_1+1)f_x.\ee Equation
(\ref{b179}) is reduced  by the substitution $y(f)=f_x$ to the form
\be\label{b180}f_x=\frac{\lbd^2(A_1+1)}{2(2A_1-1)}\ln(\gamma(t)
f),\ee where $\gamma(t)$ is an arbitrary smooth function (the case
$A_1=\frac{1}{2}$ leads again to the contradiction $g=0$).
Substituting (\ref{b180}) into the first equation of (\ref{b166})
and taking into account (\ref{b175}), we obtain
\be\label{b181}f_t=-\frac{\lbd^2(A_1+1)^2}{2A_1-1}f\ln(\gamma f).\ee

Differentiating (\ref{b180}) by $t$ and (\ref{b181}) by $x$ and
equaling  the expressions obtained, we have
\be\label{b182}\frac{\lbd^2(A_1+1)}{2A_1-1}\lf(\frac{\gamma_t}{2\gamma}+\frac{\lbd^2(A_1+1)^2}{2(2A_1-1)}(\ln(\gamma
f))^2\rg)=0.\ee The expression (\ref{b182}) is satisfying only with
$A_1=-1$, but this leads to  $f_x=0$ (see (\ref{b180})) and
therefore  $g=0$ (see (\ref{b175})). Thus,
 the
contradiction is again obtained.

Let us consider (\ref{b173}) with $A=0$. Substituting  (\ref{b173})
with $A=0$, into equation (\ref{b167}) and splitting  expression
obtained by the different powers of $V$, we obtain the system
\be\label{b169}\ba\medskip g_t+2(g+\lbd_3)(g+f_x)=0,\\ \medskip 2f_x(h+\lbd_2)+g(4h+\lbd_2)+3\lbd_3h+h_t-\lbd g_x=0,\\
\medskip2h(h+\lbd_2)+2\lbd_1 f_x-\lbd
h_x-g_{xx}=0,\\ \medskip\lbd_1 h+2\lbd_0 f_x-\lbd_0
g-h_{xx}=0.\ea\ee

Finally, any solution of the nonlinear system (\ref{b166}) and
(\ref{b169}), which consist of seven equations on three functions
$f,\ g$ and $h$ generate the operator of $Q$-conditional symmetry
\be\label{b120a}Q=\p_t+f\p_x+(gV+h)\p_V\ee
for the equation
\be\label{b178}V_{xx}=V^2V_t-\lbd VV_x+\lbd_3 V^3+\lbd_2V^2+\lbd_1
V+\lbd_0.\ee

It turns out, that the overdetermined system of PDEs (\ref{b166})
and (\ref{b169}) is compactable. However, all its solutions produce
the operators of the form (\ref{b120a}), which are nothing else but
Lie symmetry operators of (\ref{b178}) obtained in \cite{ch-se-98}.
We have established this using computer algebra package Mathematica
5.0. The relevant calculations are omitted because their
awkwardness.

\textbf{The sketch of the proof   of theorem 2 is now completed.}


\begin{thebibliography}{99}


\bibitem {ch-pliu-2006}
 Cherniha R, Pliukhin O 2007
New conditional symmetries and exact solutions of  nonlinear
reaction-diffusion-convection equations. {\it J. Phys. A} {\bf
40},\, 10049-70

\bibitem {ch-pl-06ar} Cherniha R, Pliukhin O  2006 New conditional symmetries and exact solutions of nonlinear
reaction-diffusion-convection equations I. arXiv: math-ph/0612078

\bibitem {ch-pl-07ar} Cherniha R, Pliukhin O  2007 New conditional symmetries and exact solutions of nonlinear
reaction-diffusion-convection equations II. arXiv: math-ph/0706.0814

\bibitem {a-h} Arrigo D J, Hill J M 1995
 Nonclassical symmetries for nonlinear diffusion and absorption
{\it Stud. Appl.Math.} {\bf 94} 21-39

\bibitem{ch-se-06}  Cherniha R,  Serov M 2006 Symmetries, Ansaetze  and Exact Solutions of
 Nonlinear Second-order Evolution Equations with Convection Terms, II.
  Euro.\ J.\ Appl.\ Math.\ {\bf 17}   597-605

\bibitem {zh-lahno98}    Zhdanov R Z, Lahno V I 1998 Conditional
symmetry of a porous medium  equation {\it Physica~D} \textbf{122}
178-186



\bibitem{ch-se-98}  Cherniha R, Serov M  1998
Symmetries, Ans\"atze  and Exact Solutions of  Nonlinear
Second-order Evolution Equations with Convection Term {\it Euro.\
J.\ Appl.\ Math.}\ {\bf 9} 527--542



\end{thebibliography}
\end{document}